\documentstyle[aps,epsf,prl, multicol]{revtex}

\begin{document}
\title{  Decoherence in Two Bose-Einstein  Condensates}
\author{Le-Man Kuang$^{1,2}$, Zhao-Yang Tong$^2$, Zhong-Wen Ouyang$^2$,
and Hao-Sheng Zeng$^2$}
\address{$^1$CCAST (World Laboratory), P.O. Box 8730, Beijing 100080, 
People's Republic of  China \\ 
$^2$Department of Physics, Hunan Normal University, Changsha 410081,
             People's Republic of  China$^{\dagger}$}
\maketitle
\begin{abstract}
In this paper, decoherence in a system consisting of two Bose-Einstein condensates is investigated analytically. 
 It is indicated that  decoherence  can be controlled through manipulating the interaction between the system and environment.
 The influence of the decoherence  on quantum coherent atomic tunneling (AT) between two   condensates with arbitrary 
 initial states is studied in detail. Analytic expressions of the population difference (PD) and the AT current between 
 two condensates are found. It is shown that the decoherence leads to the decay of the PD and  
 the suppression of the AT current.

\noindent PACS numbers: 03.75.Fi, 42.50.Vk, 05.30.Jp, 03.65.Bz
\end{abstract}

\pacs{03.75.Fi, 42.50.Vk, 05.30.Jp, 03.65.Bz}
 
\begin{multicols}{2}
\section{Introduction}
Recently, much attention has been paid to experimental investigations [1-7] and theoretical studies [8-15] for 
 systems   consisting  of two and multi Bose condensates since such systems give rise to a fascinating possibility 
of observing a rich set of new macroscopic quantum phenomena [16-20] which do not exist in a single condensate. 
  Among important macroscopic
quantum effects is the quantum coherent atomic tunneling (AT) between
two trapped Bose condensates [15-18]. Several authors [19,20] showed that
the AT can support  macroscopic quantum self-trapping (MQST) due to the
nonlinearity of atom-atom interactions in condensates. As is well known,
no system can be completely isolated from its environment. In fact, in
current experiments on trapped Bose condensates of dilute alkali atomic
gases condensated  atoms continuously interact with non-condensate atoms
(environment).  Interactions between a quantum  system and environment cause two types of
 irreversible effects: dissipation 
and decoherence [21,22]. Mathematically, the dissipation and decoherence can be understood in the following way. 
Let $\hat{H}_S$ and  $\hat{H}_B$ be Hamiltonian of the system and environment (bath), respectively, 
  and $\hat{H}_I$ be interacting
Hamiltonian  between the system and environment. When $[\hat{H}_I, \hat{H}_S]\neq0$, which implies that the energy of the system 
is not conservative, the interaction $\hat{H}_I$  describes 
the dissipation.   When $[\hat{H}_I, \hat{H}_S]=0$, the energy of the system 
is   conservative, so the interaction $\hat{H}_I$  describes 
the decoherence. The dissipation effect, which dissipates the energy of the quantum system into the environment, 
    is characterized by the relaxation time scale $\tau_r$.
  In contrast, the decoherence effect, which can be regarded as a mechanism 
  for enforcing classical behaviors in the macroscopic realm, is much more insidious because the coherence information 
  leaks out into the environment in another time scale  $\tau_d$, which is much shorter than $\tau_r$,  as the quantum system evolves with time.  
Since macroscopic quantum phenomena in Bose condensates 
  mainly depend on $\tau_d$ rather than $\tau_r$, the discussions in present paper only focus 
  on the decoherence problem rather than  the dissipation effect.
    To study the decoherence in Bose condensates is not only of
theoretical interest but also importance from a practical point of view,
since decohering would always be present in any  Bose condensate
experiments of trapped atoms.

On the aspect of  modeling dissipation and decoherence in trapped Bose condensates, some
progress [9,23-26] has been made. In particular,  Anglin [23] derived a master
equation for a trapped Bose condensate by considering a special model of a
condensate confined in a deep but narrow spherical square-well potential.
In his model the reservoir of non-condensate atoms consists of a continuum of
unbound modes obtained by the scattering solutions of the potential
well.   Making use of the  Anglin's master equation, Ruostekoski and Walls [24] numerically 
simulated dissipative  dynamics of a Bose condensate in a
double-well potential  when the condensate
is in the atomic coherent states,  and shown that the interactions between condensate
and non-condensate atoms make the MQST decay away. For a system consisting of two trapped weakly
connected  Bose condensates, there is quantum coherent  AT between two
condensates. Questions that naturally arise are, what is the effect
of decoherence on the quantum coherent AT ? Does decoherence
increase or decrease the AT current between them?  In this paper, we
analytically study the decoherence problem in two Bose condensate, and investigate 
the influence of the   decoherence on the quantum
coherent AT between two trapped Bose condensates in terms of an exactly solvable
Hamiltonian. We will present analytic expressions of the population difference (PD) and
the AT current between two Bose condensates, and show that the decoherence
leads to the PD decay and the suppression of the AT current.

This paper is organized as follows. In Sec.II, we present an approximate analytic solution of the system 
consisting of  two Bose condensates with a tunneling coupling without decoherence. 
In Sec. III, we introduce a decoherence model and apply it to the two-condensate system. we discuss the 
 influence of decoherence the atomic tunneling.  Concluding remarks are provided  in the last section.

\section{Two Bose-Einstein condensates with tunneling coupling}

Let us consider a system of two  Bose condensates with weak nonlinear  interatomic
interactions  and the Josephson-like coupling.   Such a condensate system,
in principle, can be  produced in a double trap with two condensates
coupled by quantum  tunneling and ground collisions, or in a system with
two different  magnetic sublevels of an atom, in which case the two
species  condensates correspond two electronic states involved.  In the formalism
of the second quantization, Hamiltonian of  such a system can be written as
\begin{equation}
\hat{H}=\hat{H}_1+\hat{H}_2+\hat{H}_{int} +\hat{H}_{Jos},
\end{equation}

\begin{eqnarray}
\hat{H}_i&=&\int d{\bf x} \hat{\psi}^{\dagger}_i({\bf x})[-\frac{\hbar^2}{2m}\nabla^2 +
V_i({\bf x})\nonumber \\
& & + U_i\hat{\psi}^{\dagger}_i({\bf x})\hat{\psi}_i({\bf
x})]\hat{\psi}_i({\bf x}), (i=1,2),
\end{eqnarray}

\begin{equation}
\hat{H}_{int}=U_{12} \int d{\bf x}
\hat{\psi}^{\dagger}_1({\bf x})\hat{\psi}^{\dagger}_2({\bf x})\hat{\psi}_1({\bf x})\hat{\psi}_2({\bf x}),
\end{equation}

\begin{equation}
\hat{H}_{Jos}=\Lambda \int d{\bf x}
[\hat{\psi}^{\dagger}_1({\bf x})\hat{\psi}_2({\bf x}) + \hat{\psi}_1({\bf
x})\hat{\psi}^{\dagger}_2({\bf x})].
\end{equation}
Here $i=1,2$, $\hat{\psi}_i({\bf x})$ and $\hat{\psi}^{\dagger}_i({\bf x})$ are the
atomic field operators which annihilate and create atoms
at position ${\bf x}$, respectively. They satisfy   the  commutation
relation $[\hat{\psi}_i({\bf x}), \hat{\psi}^{\dagger}_j({\bf x}')]=\delta_{ij}\delta({\bf x}-{\bf x}')$.
$\hat{H}_1$ and $\hat{H}_2$  describe the evolution of each species in the absence of
interspecies interactions. $\hat{H}_{int}$ describes interspecies collisions. $\hat{H}_{Jos}$ is the Josephson-like tunneling
coupling term.  Atoms are confined in harmonic potentials $V_i({\bf x}) (i=1,2)$.
 Interactions between atoms are described by a nolinear self-interaction term
 $U_i=4\pi\hbar^2a^{sc}_i/m$ and a term that corresponds the nonlinear interaction
 between different species $U_{12}=4\pi\hbar^2a^{sc}_{12}/m$, where    $a^{sc}_i$
 is $s$-wave scattering lengths of species $i$ and  $a^{sc}_{12}$ that between species 1 and 2.
 For simplicity, throughout this paper we set $\hbar=1$,  and assume that $a^{sc}_1=a^{sc}_2=a^{sc}$,
$V_1({\bf x})=V_2({\bf x})$.

It has been well known that the Hamiltonian (1) can reduce to a two-mode
Hamiltonian [27] by the use of the approximation of the atomic field
operators:$\hat{\psi}_i({\bf x})=\hat{a}_i\phi_{i}({\bf x})$,
where $\hat{a}_i=\int d{\bf x}\phi_{i}({\bf x})\hat{\psi}_i({\bf x})$
are correspondent mode annihilation operators  with real  distribution functions
$\phi_{i}({\bf x})$ and $[\hat{a}_i, \hat{a}^{\dagger}_i]=1$.
Then the Hamiltonian (1) can be reduced to  the   two-mode  Hamiltonian
\begin{eqnarray}
\hat{H}&=&\omega_0(\hat{a}^{\dagger}_1\hat{a}_1 + \hat{a}^{\dagger}_2\hat{a}_2)
+  q(\hat{a}^{\dagger2}_1\hat{a}^2_1 + \hat{a}^{\dagger2}_2\hat{a}^2_2) \nonumber \\
&& +  g(\hat{a}^{\dagger}_1\hat{a}_2 + \hat{a}^{\dagger}_2\hat{a}_1) + 2\chi\hat{a}^{\dagger}_1\hat{a}_1
\hat{a}^{\dagger}_2\hat{a}_2,
\end{eqnarray}
where $q$, $\chi$ and $g$ are coupling constants which characterize the
strength of interatomic interaction in each condensate, interspecies
interaction, and the Josephson-like coupling, respectively.

The valid conditions of the two-mode approximation were demonstrated in Ref.[19] which 
indicate that the two-mode approximation is valid for weak many-body interactions, i.e., 
for   small number of  condensated atoms. As shown in Ref.[19], the two-mode approximation
 should be acceptable for the number of atoms $N\le 2000$ if the scattering length typically taken as 
 $a=5nm$   for  a large trap with the size $L=10\mu m$. 

We note that the   two-mode approximate Hamiltonian has the same form of that of a two-mode
nonlinear optical directional coupler [28]. It can not be exactly solved, but a
closed analytical solution can be
obtained under the rotating wave approximation suggested by Alodjanc
{\it et al.} [29]. The approximate analytic solution is valid for the
weak interactions between atoms, but it sheds considerable light on
 the AT under our consideration.

In order to obtain  an approximate analytic solution of the Hamiltonian
(5), we introduce a new  pair of  bosonic operators:
\begin{equation}
\hat{A}_1=\frac{1}{\sqrt2}(\hat{a}_1 + \hat{a}_2)e^{-igt},
\hat{A}_2=\frac{i}{\sqrt2}(\hat{a}_1 - \hat{a}_2)e^{igt},
\end{equation}
 which satisfy the usual bosonic commutation relation:
$[\hat{A}_i, \hat{A}^{\dagger}_j]=\delta_{ij}$.
Then the Hamiltonian (5) reduces to the following form
\begin{eqnarray}
\hat{H}&=&\Omega\hat{N} +
g(\hat{A}^{\dagger}_1\hat{A}_1-\hat{A}^{\dagger}_2\hat{A}_2)
+ \frac{1}{4} q[(3\hat{N}^2-2\hat{N}) \nonumber \\
&& -(\hat{A}^{\dagger}_1\hat{A}_1-\hat{A}^{\dagger}_2\hat{A}_2)^2] +
\frac{1}{2}\chi\hat{N}^2
\nonumber \\
&&-\chi\hat{A}^{\dagger}_1\hat{A}_1\hat{A}^{\dagger}_2\hat{A}_2 +\hat{H}',
\end{eqnarray}
 where $\Omega=(\omega_0-\frac{\chi}{2})$, the total number operator
$\hat{N}=\hat{a}^{\dagger}_1\hat{a}_1+\hat{a}^{\dagger}_2\hat{a}_2=\hat{A}^{\dagger}_1\hat{A}_1+\hat{A}^{\dagger}_2\hat{A}_2$
is a conservative constant, and $\hat{H}'$ is a  nonresonant term   which oscillates   at the frequency $4g$ in the sense of
 the Alodjanc {\it et al.}'s proposal [29]. The account of the fast oscillating terms results only in some 
 additional oscillations which play no essential role in the evolution of the measurable quantities  
 specifying the macroscopic quantum phenomena of the two-condensate system, so that the  nonresonant 
  term are fully  negligible. This is the  rotating wave approximation (RWA)in the sense of  Ref.[29].
  After neglecting the nonresonant term   $H'$,  we get the following approximate Hamiltonian:
\begin{eqnarray}
\hat{H}_A&=&\Omega\hat{N} +
g(\hat{A}^{\dagger}_1\hat{A}_1-\hat{A}^{\dagger}_2\hat{A}_2)
+ \frac{1}{4} q[(3\hat{N}^2-2\hat{N}) \nonumber \\
&& -(\hat{A}^{\dagger}_1\hat{A}_1-\hat{A}^{\dagger}_2\hat{A}_2)^2] +
\frac{1}{2}\chi\hat{N}^2
 -\chi\hat{A}^{\dagger}_1\hat{A}_1\hat{A}^{\dagger}_2\hat{A}_2.
\end{eqnarray}

It is worthwhile noting that the dynamics of the non-RWA Hamiltonian (5) is often chaotic. A detailed investigation 
on chaotic behaviors in the two-condensate system is beyond the scope of present paper, and will be given elsewhere.
Nevertheless, the RWA Hamiltonian (8) is an integrable Hamiltonian whose dynamics is regular, does not exhibit chaos.
Hence, the the terms neglected in the RWA lead to chaos when they kept in the two-condensate system. 
This is very analogous to the case of the Jaynes-Cummings Model (JCM) [30] which describes the interaction of a 
two-level atom  with a single-mode electromagnetic field  in quantum optics. It was well known that the RWA JCM is 
exactly solvable, but the non-RWA JCM [31] exhibits chaos. As shown in Ref.[31], the chaos was a consequence of 
inclusion of terms normally neglected in the RWA.

Obviously, the Hamiltonian  $\hat{H}_A$ is diagonal in the Fock space of the
($\hat{A}_1$,
$\hat{A_2}$) representation defined by
\begin{equation}
|n,m)=\frac{1}{\sqrt{n!m!}}\hat{A}^{\dagger n}_1\hat{A}^{\dagger m}_2|0,0),
\end{equation}
where $n$ and $m$ take nonnegative integers. And we have $\hat{H}_A|n,m)=E(n,m)|n,m)$ with the eigenvalues
\begin{eqnarray}
E(n,m)&=&(\Omega-\frac{q}{2})(n+m) + g(n-m) + \frac{1}{4}(3q+2\chi)\nonumber
\\
&&\times (n+m)^2
       - \frac{q}{4}(n-m)^2 - \chi nm.
\end{eqnarray}

For simplicity, all calculations below shall be carried out  in the ($\hat{A}_1$,
$\hat{A}_2$) representation  with the basis: $\{|n,m), n,m=0,1,2,...\}$, which is
related to  the ($\hat{a}_1$, $\hat{a}_2$) representation with
a set of   basis: $\{|n,m\rangle =\hat{a}^{\dagger n}_1\hat{a}
^{\dagger m}_2/\sqrt{n!m!}|0,0\rangle,
n,m=0,1,2,...\}$ through
the following relation
\begin{eqnarray}
|n,m\rangle &=&\sum^n_{r=0}\sum^m_{s=0}\frac{[n!m!(n-r+s)!(m-s+r)!]^{1/2}}
{2^{(n+m)/2}(n-r)!(m-s)!}
 \nonumber \\
& &\times e^{-i(2r-3m)\pi/2}|n-r+s,m-s+r).
\end{eqnarray}

\section{Influence of decoherence on  atomic tunneling}

We now consider the effect of the decoherence.
 We use a reservoir consisting of an infinite set of
harmonic oscillators to model environment of condensate atoms in a trap, and
we assume the total Hamiltonian  to be
\begin{eqnarray}
\hat{H}_T&=&\hat{H}_A + \sum_k\omega_k\hat{b}^{\dagger}_k\hat{b}_k
+ F(\{\hat{S}\})\sum_kc_k(\hat{b}^{\dagger}_k+\hat{b}_k)
\nonumber \\
&&+F(\{\hat{S}\})^2\sum_k\frac{c_k^2}{\omega_k^2},
\end{eqnarray}
where the second term is the Hamiltonian of the reservoir. The last term in Eq.(12)  is a renormalization term
[32].  The third term in Eq.(12) represents the interaction between the system and the reservior with
a coupling constant $c_k$, where $\{\hat{S}\}$ is a set of linear operators 
of the system or their linear combinations in the same picture as that of $\hat{H}_A$, 
$F(\{\hat{S}\})$ is an operator function  of $\{\hat{S}\}$. In order to  enable  
 what the interaction between the system and  environment describes is decoherence  not dissipation, 
 we require that the linear operator $\hat{S}$  commutes with the the Hamiltonian of the system $\hat{H}_A$.  
 Then, the interaction term in Eq.(12) commutes with the Hamiltonian of the system. This implies  that there is 
 no   energy transfer between the system and its environment. So that  it  does describe  the decoherence. 
 The concrete form of the function $F(\{\hat{S}\})$, which may be considered  as  an experimentally determined quantity, 
 may be   different   for different environment. Therefore, the decohering interaction 
in (12) can not only describe decoherence caused by  the effect of elastic collisions between condensate and 
non-condensate atoms for a Bose condensate system, but also simulate decoherence caused by other decoherencing sources 
through properly  choosing  the operator function of the system $F(\{\hat{S}\})$. 
 
The Hamiltonian (12) can be exactly solved by making use of the following
unitary transformation
\begin{equation}
\hat{U}=\exp[\hat{H}_A\sum_k\frac{c_k}{\omega_k}(\hat{b}^{\dagger}_k-\hat{b}_k)].
\end{equation}

Corresponding to the Hamiltonian (12), the total density operator of the
system plus reservoir can be expressed as
\begin{eqnarray}
\hat{\rho}_T(t)&=&e^{-i\hat{H}_At}\hat{U}^{-1}e^{-it\sum_k\omega_k\hat{b}^{\dagger}_k\hat{b}_k}\hat{U}
\hat{\rho}_T(0)\hat{U}^{-1}\nonumber
\\
&&\times e^{it\sum_k\omega_k\hat{b}^{\dagger}_k\hat{b}_k}\hat{U}e^{i\hat{H}_At}.
\end{eqnarray}
In the derivation of the above solution, we have used
$\hat{\rho}_T(t)=\hat{U}^{-1}\hat{\rho}'_T(t)\hat{U}$, where
$\hat{\rho}'_T=e^{-i\hat{H}'_Tt}\hat{\rho}'_T(0)e^{i\hat{H}'_Tt}$
with $\hat{H}'_T=\hat{U}\hat{H}_T\hat{U}^{-1}$ and
$\hat{\rho}'_T(0)=\hat{U}\hat{\rho}_T(0)\hat{U}^{-1}$, where $\hat{\rho}_T(0)$
the initial total density operator.

We assume that the system and  reservoir are initially in thermal
equilibrium and  uncorrelated, so that
$\hat{\rho}_T(0)=\hat{\rho}(0)\otimes\hat{\rho}_R$, where
$\hat{\rho}(0)$ is the initial  density operator of the system,
and $\hat{\rho}_R$ the density operator of the reservoir, which can be
written as $\hat{\rho}_R=\prod_k\hat{\rho}_k(0)$ with $\hat{\rho}_k(0)$
is the density  operator of the $k$-th harmonic oscillator in thermal
equilibrium. After taking the  trace over the
reservoir, from  Eq.(14) we can get the reduced density
operator of the system, denoted by $\hat{\rho}(t)=tr_R\hat{\rho}_T(t)$,
its matrix elements in the ($\hat{A}_1$, $\hat{A}_2$) representation
 are explicitly written as
\begin{eqnarray}
\rho_{(m',n')(m,n)}(t)&=&\rho_{(m',n')(m,n)}(0)R_{(m',n')(m,n)}(t)
\nonumber \\
&&\times e^{-i[F(\{S(m',n')\})-F(\{S(m,n)\})]t},
\end{eqnarray}
where $F(\{S(m,n)\})$ is an eigenvalue of the operator function $F(\{\hat{S}\})$ in 
an eigenstate of $\hat{H}_A$.
 $R_{(m',n')(m,n)}(t)$ is a reservoir-dependent quantity given by
\begin{eqnarray}
R_{(m',n')(m,n)}(t)&=&
e^{-i[F^2(\{S(m',n')\})-F^2(\{S(m,n)\})]Q_1(t)}
\nonumber \\
& &\times e^{-[F(\{S(m',n')\})-F(\{S(m,n)\})]^2Q_2(t)},
\end{eqnarray}
where  the two reservoir-dependent functions are given by
\begin{equation}
Q_1(t)=\int^{\infty}_{0} d\omega J(\omega)\frac{c^2(\omega)}{\omega^2}\sin(\omega t),
\end{equation}

\begin{equation}
Q_2(t)=2\int^{\infty}_{0} d\omega J(\omega)
\frac{c^2(\omega)}{\omega^2}\sin^2(\frac{\omega t}{2})\coth(\frac{\beta\omega}{2}),
\end{equation}
Here we have taken the continuum limit of the reservoir modes: $\sum_k \rightarrow
\int^{\infty}_{0} d\omega J(\omega)$,  where $J(\omega)$ is the spectral density of
the reservoir, $c(\omega)$ is the correspondeing continuum expression for $c_k$, and
$\beta=1/k_BT$ with $k_B$ and $T$ being  the Boltzmann constant
and temperature, respectively.

It is well known that decoherence corresponds to the decay of off-diagonal  elements 
of the reduced density matrix of a quantum system. For the case under our consideration, 
the degree of decoherence is determined by the decaying factor in Eq.(16). 
It is interesting  to note that if we choose a proper operator function $F(\{\hat{S}\})$ 
to make $F(\{S(m',n')\})=F(\{S(m,n)\})$ for $(m',n')\neq (m,n)$, then we find that
\begin{equation}
\rho_{(m',n')(m,n)}(t)=\rho_{(m',n')(m,n)}(0)
\end{equation}
 which indicates that  the quantum system maintains its initial quantum coherence, 
 namely, the time evolution of decoherence-free of the quantum system is realized. Therefore, 
 we conclude that one can control decoherence by manipulating interaction function  $F(\{\hat{S}\})$.

Eqs.(15) and (16) indicate that the interaction between the system and
its environment induces a phase shift and a decaying factor in  the
reduced density operator of the system. We now consider the PD
 between the two condensates in the presence of the decoherence,
defined by  $p(t)\equiv N_1(t)-N_2(t)$ with $N_i=\langle
\hat{a}^{\dagger}_i\hat{a}_i\rangle$. We find that  
\begin{eqnarray}
p(t)&=&-2\sum_r\sum_s\sqrt{s(r+1)}|\rho_{(r+1,s-1)(r,s)}(0)|
 \nonumber \\
&&\times \sin[\theta_{rs}-v^-_{rs}(t-v^+_{rs}Q_1(t))]e^{-(v^-_{rs})^2Q_2(t)},
\end{eqnarray}
where we have introduced  the symbols: 
\begin{equation}
\rho_{(r+1,s-1)(r,s)}(0)=|\rho_{(r+1,s-1)(r,s)}(0)|e^{i\theta_{rs}},
\end{equation}

\begin{equation}
v^{\pm}_{rs}=F(\{S(r+1,s-1)\})\pm F(\{S(r,s)\}).
\end{equation}

From Eq.(20) we see that if we do not take into account the influence  of
the decoherence, i.e., set $Q_1(t)=Q_2(t)=0$, then  we get a expression of
the PD  between two condensates 
\begin{eqnarray}
p(t)&=&-2\sum_r\sum_s\sqrt{s(r+1)}|\rho_{(r+1,s-1)(r,s)}(0)|
\nonumber \\ 
&&\times\sin(\theta_{rs}-v^-_{rs}t),
\end{eqnarray}
 which implies that the time evolution of the PD is periodic. In particular, if we 
 take  $F(\{\hat{S}\}=\hat{H}_A$, we find that when   $2g/(q-\chi)=K$ ( being an integer), 
 we have a nonzero time-average value of the PD
\begin{eqnarray}
\bar{p}&=&-2\sum_r\sqrt{(r-K+1)(r+1)}|\rho_{(r+1,r-K)(r,r-K+1)}(0)|
\nonumber \\
&&\times \sin(\theta_{r r-K+1}),
\end{eqnarray}
which means that the two-condensate system under our consideration
exhibits the MQST [18,19] when the   decoherence is absent.

The  coherent AT current between the two condensates,
 defined by $I(t)\equiv\dot{N}_1(t)-\dot{N}_2(t)$,  is given by
\begin{eqnarray}
I(t)&=&2\sum_r\sum_s\sqrt{s(r+1)}v^-_{rs}|\rho_{(r+1,s-1)(r,s)}(0)|
\nonumber \\
& &\times \{(1-v^+_{rs}\dot{Q}_1(t))\cos[\theta_{rs}-v^-_{rs}(t-v^+_{rs}Q_1(t))]
\nonumber \\
& &+ v^-_{rs}\dot{Q}_2(t)\sin[\theta_{rs}-v^-_{rs}(t-v^+_{rs}Q_1(t))]\}
\nonumber \\
& &\times e^{-(v^-_{rs})^2Q_2(t)}.
\end{eqnarray}

From Eqs.(18), (20) and (25) we can immediately draw one important qualitative
conclusion:  since $Q_2(t)$ is positive definite, the existence of the
decoherence is always to tend to suppress the PD and the AT
current between the two condensates. This answers the question: ``Does the
decoherence increase or decrease the AT?".

From Eqs.(17),(18), (20) and (25) we see that all necessary information about the
effects of the environment on the PD  and the AT current is
contained in the spectral density of the reservoir. To procced further
let us now specialize to the Ohmic  case [33] with  the  spectral distribution
$J(\omega)=\frac{\eta\omega}{c^2(\omega)}e^{-\omega/\omega_c}$, where $\omega_c$ is
the high frequency cut-off, $\eta$ is a positive characteristic parameter of
the reservoir. With this choice, at low temperature the functions $Q_1(t)$
and  $Q_2(t)$ are given by the following expressions
\begin{equation}
Q_1(t)=\eta\tan^{-1}(\omega_c t),
\end{equation}

\begin{equation}
Q_2(t)=\eta\{\frac{1}{2}\ln[1+(\omega_ct)^2] + \ln[\frac{\beta}{\pi
t}\sinh(\frac{\pi t}{\beta})]\}.
\end{equation}

Recent experiments [1,7] on two condensates  have established a typical time scale at 
which the two condensates preserve coherence. The value of the typical time scale is $t\dot{=}100ms$.
In the meaningful  domain of time $\omega_ct\gg 1$ which requires $\omega_c\gg 10Hz$ which can be easily 
satisfied  for an usual reservoir [34,32,33],    at zero temperature, we have   
$\dot{Q}_1(t)\doteq \eta /(\omega_ct^2)$, and $Q_2(t)\doteq \eta\ln(\omega_ct)$, then we find
\begin{eqnarray}
p(t)&=&-2\sum_r\sum_s\sqrt{s(r+1)}|\rho_{(r+1,s-1)(r,s)}(0)|
\nonumber \\
& &\times \sin[\theta_{rs}-v^-_{rs}(t-v^+_{rs}Q_1(t))](\omega_ct)^{-\eta
(v^-_{rs})^2},
\end{eqnarray}

\begin{eqnarray}
I(t)&=&2\sum_r\sum_s\sqrt{s(r+1)}v^-_{rs}|\rho_{(r+1,s-1)(r,s)}(0)|
\nonumber \\
& &\times \{(1-\frac{\eta
v^+_{rs}}{\omega_c}t^{-2})\cos[\theta_{rs}-v^-_{rs}(t-v^+_{rs}Q_1(t))]
\nonumber \\
& &+ \eta v^-_{rs}t^{-1}\sin[\theta_{rs}-v^-_{rs}(t-v^+_{rs}Q_1(t))]\}
\nonumber \\
& &\times (\omega_ct)^{-\eta (v^-_{rs})^2},
\end{eqnarray}
which indicate that the PD and the AT current decay
away according to the ``power law", where we have noted that the decaying
factors can not be taken outside the summation on the r.h.s. of Eqs.(28) and
(29).

At finite temperature, we have $\dot{Q}_1(t)\doteq \eta
/(\omega_ct^2)$, and $Q_2(t)\doteq
\eta[\ln(\frac{\beta\omega_c}{2\pi})+ \frac{\pi t}{\beta}]$, so that

\begin{eqnarray}
p(t)&=&-2\sum_r\sum_s\sqrt{s(r+1)}|\rho_{(r+1,s-1)(r,s)}(0)|
\nonumber \\
& &\times \sin[\theta_{rs}-v^-_{rs}(t-v^+_{rs}Q_1(t))]
 \nonumber \\
& &\times (\frac{\beta\omega_c}{2\pi})^{-\eta  (v^-_{rs})^2 }
\exp[-\frac{\eta (v^-_{rs})^2\pi}{\beta}t],
\end{eqnarray}

\begin{eqnarray}
I(t)&=&2\sum_r\sum_s\sqrt{s(r+1)}v^-_{rs}|\rho_{(r+1,s-1)(r,s)}(0)|
\nonumber \\
& &\times \{(1-\frac{\eta
v^+_{rs}}{\omega_c}t^{-2})\cos[\theta_{rs}-v^-_{rs}(t-v^+_{rs}Q_1(t))]
\nonumber \\
& &+ \frac{\eta \pi
v^-_{rs}}{\beta}\sin[\theta_{rs}-v^-_{rs}(t-v^+_{rs}Q_1(t))]\}
\nonumber \\
& &\times (\frac{\beta\omega_c}{2\pi})^{-\eta  (v^-_{rs})^2 }
\exp[-\frac{\eta (v^-_{rs})^2\pi}{\beta}t],
\end{eqnarray}
which indicate that at finite temperature  the PD  decays
away according to the ``exponential law", and the decay of the AT current
becomes  more complicated than that of the PD
 due to the factor $(\eta v^+_{rs}/\omega_c)t^{-2}$ before the
cosine function in Eq.(31).

\section{Concluding remarks}

 We have present a decoherence model which is exactly solvable, and applied it to study 
decoherence in two Bose condensates. We have indicated that one can control
 decoherence by manipulating interaction between a quantum system and environment.
 We have investigated the influence of the decoherence
 on quantum coherent AT  between two trapped  Bose  condensates with
 arbitrary initial states, and shown that the decoherence suppresses
the PD and the AT current between two condensates. We have obtained analytic
expressions of the PD and the AT current and found  that for the
 reservoir-spectral density of the  Ohmic case, the PD
and the AT current decay away by the ``power law" at zero temperature; at finite
temperature, the PD decays away by the ``exponential law"  while the decay of
the AT current contains both  ``exponential-law" and ``power-law" components.
It is worthwhile to note that
our results are obtained for arbitrary initial states of the two condensates,
our entire analysis is carried out without invoking the assumption of Bose-broken
symmetry which has recently  been shown to be unnecessary for a Bose condensate of
trapped atoms [35].   Also it should be  pointed out that these results are
obtained under the RWA in the sense of Alodjanic {\it et al.}'s proposal, they are valid for
interatomic weak nonlinear interactions in two condensates. The RWA essentially changes the two-condensate 
system into an integrable system, hence it suppresses the chaotic behaviors of the two-condensate 
system.

 \begin{center}
 {ACKNOWLEDGMENTS}
 \end{center}
 
L. M. Kuang would like to  acknowledge the  Abdus Salam
International Center for Theoretical Physics, Trieste, Italy
for its hospitality where part of this work was done.
This work was supported in part by the climbing project of China, NSF of China, NSF of Hunan
Province,  special project of NSF of China via Institute of Theoretical Physics,
Academia Sinica.

\end{multicols}
\end{document}